\documentclass[11pt]{article}
\usepackage{moriond,epsfig}

\begin{document}
\vspace*{4cm}
\title{ON STOKES POLARIMETERS FOR HIGH PRECISION CMB MEASUREMENTS AND MM ASTRONOMY MEASUREMENTS}

\author{M. SALATINO and P. DE BERNARDIS }

\address{Department of Physics University of Rome ``la Sapienza'', p.le Aldo Moro, 2\\
00185, Rome, Italy}

\maketitle\abstracts{Several on-going and future  experiments use
a Stokes polarimeter (i.e. a rotating wave plate followed by a
steady polarizer and by an unpolarized detector) to measure the
small polarized component of the Cosmic Microwave Background. The
expected signal is typically evaluated using the Mueller
formalism. In this work we carry-out the signal evaluation taking
into account the temperatures of the different optical devices
present in the instrument, their non-idealities, multiple internal
reflections, and reflections between different optical components.
This analysis, which exploits a new description of the radiation
transmitted by a half wave plate, can be used to optimize the
experimental setup as well as each of its optical components. We
conclude with an example of application of our analysis, studying
a cryogenic polarization modulator developed for detecting the
interstellar dust polarization.}

\section{Introduction}
The most ambitious challenge in Experimental Cosmology today is the
precision measurement of the polarized signal of the Cosmic
Microwave Background (CMB). Given the tiny amplitude of the
polarized component with respect to the unpolarized one, its
extraction is very complicated. Several on-going and future CMB
experiments, like EBEX \cite{Oxley04}, BRAIN \cite{Polenta07},
BICEP-2 \cite{Kova07}, QUBIC \cite{Kapl09}, SPIDER \cite{Crill08},
LSPE \cite{deBe09}, B-Pol \cite{deBernardis09}, CMB-Pol
\cite{Bock09}, and astrophysical ones, like PILOT \cite{Bernard07}
and BLAST-pol \cite{Mars08}, will detect the polarized component of
the signal of interest (CMB or interstellar dust polarization) by
means of a rotating a Half Wave Plate (HWP) followed by a fixed
polarizer, both located in front of the detector.

Using the Stokes parameters and the Mueller matrix formalism, the
power detected when an ideal HWP rotates at angular speed $\omega$
in front of an ideal polarizer is \cite{Collett93}:
\begin{equation}
S_{ideal}(\theta)=\frac{1}{2}[S_0+S_1 \cos{4\theta}+S_2\sin{4\theta}];
\end{equation}
where $\theta=\omega t$ and $S_i$ are the Stokes parameters of the
radiation under analysis. The linearly polarized radiation is thus
modulated at frequency $4f=2\omega/\pi$. Only the radiation
transmitted by the HWP + polarizer stack have been included. In
general, we need to include in the analysis also the thermal
emission from all the optical devices and from the radiative
background. Moreover the non-ideality of the devices, the multiple
internal reflections and the reflections between distinct devices
modify the signal detected by the polarimeter. In
Sec.$\,$\ref{sec:theory} we describe the behavior of a real
polarimeter. We provide simulations of the signal detected by a
typical CMB experiment in Sec.$\,$\ref{sec:appl}. Before concluding
summarizing our results (Sec.$\,$\ref{sec:concl}), we present the
Cryogenic Waveplate Rotator (CWR), a system which will modulate the
polarized component of the interstellar dust in the PILOT
balloon-borne experiment (Sec.$\,$\ref{sec:CWR}).

\section{Modeling a real polarimeter}\label{sec:theory}

The Mueller formalism as used above provides a simplistic
description of the radiation transmitted by an ideal HWP; in
particular, it assumes that the 100$\,\%$ of the incident
radiation is transmitted, independently of the incidence angle,
and that the phase difference $\phi$ is frequency-independent. In
the Adachi formalism \cite{Adachi60}, instead, the phase
difference depends on the frequency of the incident wave, on the
thickness of the crystal $d$, on the extraordinary $n_e$ and
ordinary $n_o$ refraction index of the birefringent crystal.

Here we use a new description of the HWP \cite{Salatino10a} that
takes into account multiple internal reflections. These depend on
the optically activity of the anisotropic HWP crystal (OAMR,
Optically Active Multiple Reflections), and we also include the
frequency dependence of the refraction indices. From the total
transmitted and reflected electric fields we have built two new
Mueller matrices: one for the transmitted stokes vector and the
other for the reflected one; the matrix elements depend on $n_e$,
$n_o$, $\phi$, the reflectivity, and the transmissivity of the
anisotropic medium in a complex way. For a real HWP,
$M_{HWP}^{real}$, in normal incidence approximation the multiple
reflections create non-vanishing off-diagonals terms, not present in
the ideal one, $diag(M_{HWP}^{ideal})=(1,1,-1,-1)$. For example, at
150$\,$GHz, a typical Mueller matrix for the HWP is:
\begin{equation} M_{HWP}^{real}=\left(
  \begin{array}{cccc}
    0.773 & -0.006 &  0     &  0     \\
   -0.006 &  0.773 &  0     &  0     \\
    0     &  0     & -0.773 & -0.033 \\
    0     &  0     &  0.033 & -0.773 \\
  \end{array}
\right).
\end{equation}

\begin{figure}
\centerline{\psfig{figure=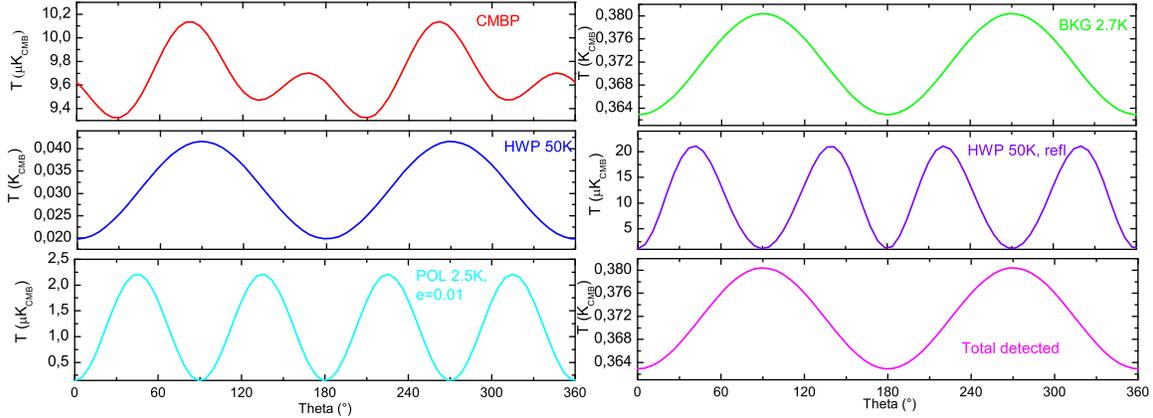,height=7.5cm}}
\caption{Simulation of the different contributions to the power
detected in a Stokes polarimeter at 150$\,$GHz, vs rotation angle of
the waveplate. From top left to bottom right: CMB polarization,
unpolarized radiative background with temperature 2.7$\,$K,
polarized emission of the HWP at 50$\,$K transmitted to the
polarizer and the same reflected back by the polarizer, polarized
emission of the polarizer ($T_p=2.5K$, $\epsilon_p= 0.01$), total
signal detected by the bolometer.}\label{fig:1}
\end{figure}

We also take into account the spectral dependence of the refraction
indices and of the absorption coefficients which are assumed to
decrease linearly with the temperature of the HWP.

A polarizer emits linearly polarized radiation; in the Stokes
polarimeter this emission is reflected back by the rotating HWP, and
after the crossing through the polarizer, is modulated at a
frequency 4$f$. The unpolarized radiative background, following the
same optical path of the astrophysical signal, is instead modulated
at 2$f$ by small non-idealities in the wave plate. Small differences
in the absorbtion coefficients of the HWP, fews $10^{-3}$, produce a
polarized emission; it is modulated at 2$f$ (4$f$) when is
transmitted (reflected back) by a polarizer (and successively by the
HWP) \cite{Salatino10a}.

\begin{figure}
\centerline{\psfig{figure=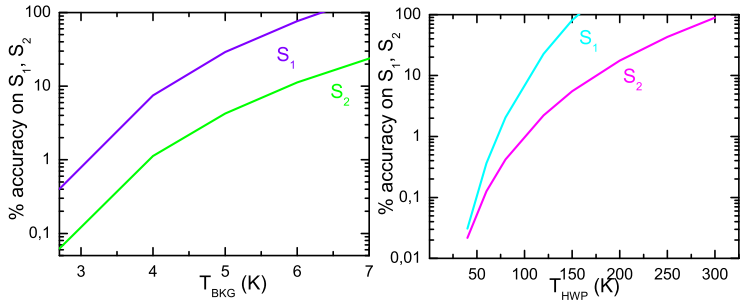,height=5.5cm}} \caption{Accuracy
for the detection of linear CMB polarization versus the temperature
of the radiative background (left) and of the HWP
(right).\label{fig:2}}
\end{figure}

\begin{table}[t]
\caption{Features of the spurious signals produced by a real
polarimeter. \label{tab:est}} \vspace{0.4cm}
\begin{center}
\begin{tabular}{|c|c|c|}
\hline
EFFECT & MOD. & AMPL.\\
\hline radiative background at 2.7$\,$K                                        & 2$f$& 15$\,$mK  \\
polarized emission of the polarizer at 2.5$\,$K reflected by the HWP
& 4$f$ & 2.1$\,\mu$K \\
polarized emission of the HWP at 50$\,$K transmitted by the
polarizer & 2$f$ & 22$\,$mK \\
polarized emission of the HWP at 50$\,$K reflected by the polarizer
& 4$f$& 3$\,\mu$K \\
polarized emission of the HWP at 4$\,$K reflected by the polarizer
& 4$f$& 0.6$\,\mu$K \\
\hline
\end{tabular}
\end{center}
\end{table}

For a $\Lambda$-CDM model, with tensor to scalar ratio $r=0.1$, the
expected detected signal due only to the CMB, has a typical (rms)
amplitude  of about $0.7\mu\,$K. We consider the normal incidence
approximation and an incoming monochromatic ray at 150$\,$GHz. Going
through a real absorbing HWP the signal amplitude does not change
much (0.6$\,\mu$K), but the heights of the peaks become uneven, due
to a spurious 2$f$ component (Fig.$\,$\ref{fig:1}). From our model,
we find a number of spurious effects, as quantified in
Tab.$\,$\ref{tab:est}.

The result of the sum of these emissions, with respect to the CMB
signal, depends on the relative angle between the CMB polarization
angle and the orientation of the wire grid. Despite of the large
amplitude of the 2$\theta$ components, they are easily removable
from the cosmological signal by means of a high-pass filter. The
4$f$ spurious signals, on the contrary, cannot be removed; being
at the same frequency, they contaminate the cosmological signal by
an amount which depends on the physical temperature of the optical
components: at 2.5$\,$K the spurious signal is about 4 times the
CMB one, while at 1$\,$K it decreases below 4$\%$ of the CMB one.
We conclude that it is certainly necessary to cool down the
polarizer and the HWP, and to reduce the radiative background.

\section{Saturation effects}\label{sec:appl}

Bolometric detectors, used at these frequencies, are non linear and
start to saturate if the detected signals become too large. In a
slightly saturated bolometer a pure 2$\theta$ signal acquires a
4$\theta$ component, plus higher order terms. Therefore, non
linearities place upper limits on the radiative background we can
tolerate in a CMB polarization experiment. Given a 1$\%$ saturation
for signals of the order of 200$\,$m$K_{CMB}$, we find that if we
want to detect the $S_1$ and $S_2$ parameters with, at least, 10$\%$
accuracy, we must reduce the radiative background below 4.5$\,$K
(Fig.$\,$\ref{fig:2}).

A possible solution for removing these troublesome spurious signals
could be using an array of polarizers, one per detector, with
different wire grid orientations, in place of a single wire grid
covering all the array. In this way, the spurious emissions having
different phases partially cancel each other. PSBs are naturally
providing this advantage, if properly oriented in the array.

\section{The Cryogenic Waveplate Rotator}\label{sec:CWR}

In the near future PILOT \cite{Bernard07}, a balloon experiment
with 1024 bolometers cooled down to 0.3$\,$K, will observe the
polarized emission from the Galactic plane and the cirrus clouds
at high Galactic latitudes, in two bands centered on 545 and
1250$\,$GHz. The cryogenic waveplate rotator (CWR)
\cite{Salatino08} \cite{Salatino10b} will modulate the
astrophysical signal at a frequency of 4$f$, rotating a 4K HWP.
The CWR works in a step and integration mode, exploiting an
innovative mechanical system driven by a step motor running at
room temperature. The CWR controls, in a completely automated way,
the position of the crystal with an accuracy better than
0.03$^{\circ}$, stopping in pre-selected positions sensed by a
3-bit optical encoder made with optical fibers.

\section{Conclusions}\label{sec:concl}

We have shown a few examples of contamination of Stokes Polarimeter
data due to the polarized emission of the polarizer reflected back
by the rotating HWP, to the unpolarized radiative background, which
follows the same path of the target signal, going through a
non-ideal waveplate, and to the polarized emission of the HWP
reflected and transmitted by a polarizer. This study has used a new
description of the HWP which takes into account the optically active
multiple reflections inside the crystal and which describes the
radiation transmitted and reflected from the birefringent medium.
The non linear behavior of the bolometric detector produces a
4$\theta$ signal from the large 2$f$ spurious signals; this means
that even if the 2$f$ signal can be removed post-detection, its
level should be kept low enough that the linearity of the detector
is not challenged. This sets upper limits, less than about 10$\,$K,
for the allowed radiative background in a given experiment. Our
simulations also show the necessity of cooling the polarizer down to
about 1$\,$K and the HWP to 4$\,$K.

\end{document}